\documentclass[preprint,5p,times]{elsarticle}

\usepackage{amsmath,amssymb,amsfonts,amsgen}

\usepackage{graphicx}
\usepackage{array}
\usepackage{float}
\usepackage{hyperref}
\usepackage{dcolumn}
\usepackage{bm}
\usepackage{cases}
\usepackage{color}
\usepackage{comment}

\newcommand{\gerris}{{\usefont{T1}{pzc}{m}{n}Gerris}}
\newcommand{\ff}{{\tt FreeFem++}}

\graphicspath{{figs/}}

\journal{European Journal of Mechanics - B/Fluids}

\begin{document}

\begin{frontmatter}

\title{A pressure impulse theory for hemispherical liquid impact problems}

\author[label1]{Julien Philippi\corref{cor1}\fnref{label2}}
\address[label1]{Sorbonne Universit\'es, UPMC Univ Paris 06, CNRS, UMR 7190 Institut Jean Le Rond d'Alembert, F-75005 Paris, France.}

\cortext[cor1]{Corresponding author.}
\fntext[label2]{Present address : Nonlinear Physical Chemistry Unit, Faculté des Sciences, Université libre de Bruxelles (ULB), CP231, 1050 Brussels, Belgium.}
\ead{Julien.Philippi@ulb.ac.be}

\author[label1]{Arnaud Antkowiak}

\author[label1]{Pierre-Yves Lagr\'ee}

\date{\today}

\begin{abstract}
Liquid impact problems for hemispherical fluid domain are considered. By using the concept of pressure impulse we show that the solution of the flow induced by the impact is reduced to the derivation of Laplace's equation in spherical coordinates with Dirichlet and Neumann boundary conditions. The structure of the flow at the impact moment is deduced from the spherical harmonics representation of the solution. In particular we show that the slip velocity has a logarithmic singularity at the contact line. The theoretical predictions are in very good agreement both qualitatively and quantitatively with the first time step of a numerical simulation with a Navier-Stokes solver named \gerris.
\end{abstract}

\begin{keyword}
Liquid impact \sep Drop \sep Pressure impulse
\end{keyword}

\end{frontmatter}


\section{Introduction}
Impacts are extremely brief and violent phenomena involving solids or fluids occurring in very diverse situations: impact of a wave on a seawall \citep{Cooker1995}, the landing of seaplanes \citep{Wagner1932} and more generally the water entry of solid objects \citep{Howison1991,Truscott2014} or the impact of a liquid drop onto a solid surface \citep{Josserand2016,Philippi2016}.
\begin{figure}
\begin{center}
\includegraphics{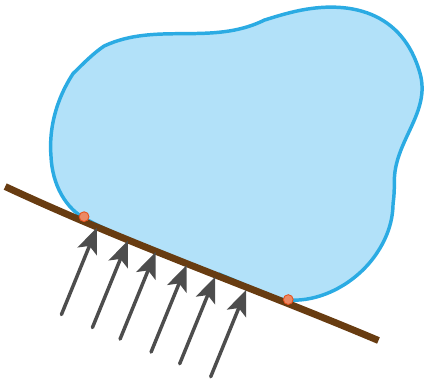}
\caption{Sketch of an impact problem involving a liquid domain of arbitrary shape and a solid boundary. The red dots represent the position of the contact line.}
\label{fig:General_Impact_Problems}
\end{center}
\end{figure}
As suggested by the variety of previous examples, impact problems involve any geometry and could be defined as a general problem with liquid and solid domains of arbitrary shape (see Fig.~\ref{fig:General_Impact_Problems}). 

The impulsive aspect of impacts is a common characteristic shared by all these problems. This feature could be defined as a considerable acceleration of a boundary of the system over a very short time. Consequently impact phenomena are unsteady, non-linear and could produce large deformations as in the case of problems involving free-surfaces \emph{e.g.} the run across a river of the Jesus-Christ lizard \citep{Hsieh2004} or the game of stone-skipping \citep{Bocquet2003}. This last problem was applied to naval artillery and studied theoretically by E. de Jonqui\`eres \citep{DeJonquieres1883} in order to explain why the bouncing of cannonball across water improves the range of the shoot.

In this paper we will focus on impact problems involving drops with the emphasis on its impulsive aspect. More specifically we will consider the particular case of hemispherical fluid domains.
\begin{figure}
\begin{center}
\includegraphics[width=9cm]{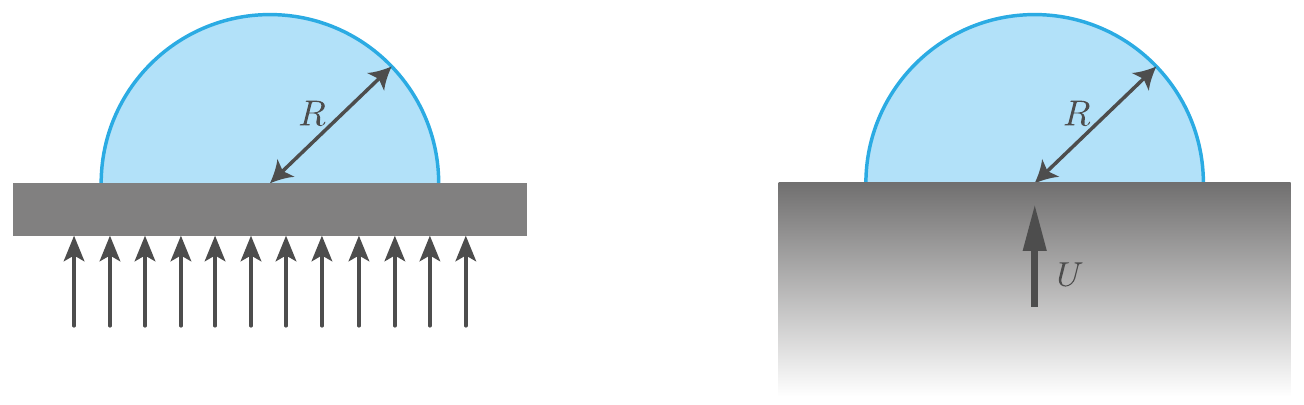}
\caption{Left: Sketch of a drop of radius $R$ sitting on a solid substrate. The latter is initially impacted from beneath. The impulsion induces deformations of the drop and eventually its partial or total ejection if the impact energy is sufficiently large. Right: Sketch of the problem studied in this paper. A drop is disposed on an nondeformable cylindrical substrate of larger radius. The latter rises impulsively with a velocity $U$ toward the drop at time $t=0$. This problem could also be seen as an impact problem by considering the free fall of a cylinder with a drop disposed at its top. The substrate rises at a velocity $U$ in the reference frame of the drop as soon as the cylinder hits the soil.}
\label{fig:DefinitionProblemes}
\end{center}
\end{figure}
One example of problem worth of interest in this kind of configuration is the study of the dynamics of a drop sitting on a solid substrate when this last is impacted from beneath (see Fig.~\ref{fig:DefinitionProblemes} left). There are several interesting questions associated to this problem such as : \emph{(i)} What is the minimal impact intensity necessary to eject partially or totally the drop from the solid surface ? \emph{(ii)} What are the deformation modes induced by the impact ? \emph{(iii)} What is the influence of the substrate's inclination ? The two first problematics have already been treated for sessile drops sitting on an oscillating solid substrate \citep{Chang2015,Wilkes2001}. The solution of the whole problem as defined here is an ambitious program clearly beyond the scope of this paper. Then we propose to solve a slightly different problem with the same impulsive characteristic. Henceforth we consider a drop of radius $R$ sitting on a larger cylindrical substrate. The substrate is initially risen impulsively toward the drop with a velocity $U$ (see Fig.~\ref{fig:DefinitionProblemes} right). This new configuration, based on the impulsive motion of a solid boundary has already been considered in different contexts \citep{Stewartson1951,Roberts1987}. 

The impulsive problem we propose here could also be seen as an impact problem. In fact it is possible to study it experimentally by using a cylinder, assumed nondeformable, with a drop disposed at its top and falling in free fall. When the cylinder hits the ground, the impact imposes a velocity $U$ of the substrate in the reference frame of the drop. Hence we can consider that these points of view are both equivalent. In terms of impulsive impact of liquid bodies on plane wall, a similar problem was studied for different geometries by Tyvand \emph{et al.} \citep{Tyvand2014} and for cylindrical fluid bodies by Hjelmervik \emph{et al.} \citep{Hjelmervik2016}. The aim of the present paper is to study impact problems for a drop disposed onto a cylindrical substrate by using the analogy with the impulsive problem depicted here and with a focus on the flow at the impact moment. In \S 2 we introduce the theoretical framework of the problem, based on the concept of pressure impulse and associated to the impulsive nature of impacts. We deduce that the problem is reduced to the solution of Laplace's equation with Dirichlet and Neumann boundary conditions. In \S 3 we determine the pressure impulse along the wetted region and the slip velocity by using a spherical harmonics representation of the solution. Then, theoretical predictions are qualitatively and quantitatively compared with numerical solutions. In \S 4 we depict the structure of the flow induced by the impact. We deduce that the slip velocity is logarithmically divergent close to the contact line. The momentum's lost due to the impact is also computed. Finally, in \S 5 we summarize our main results. The solution of several two-dimensional impact problems for planar and circular geometries is presented in appendix. Each problem is solved with a different method.

\section{Model}
\label{sec:Model}
\subsection{Pressure impulse theory}
Since the pioneering works of Wagner \cite{Wagner1932} on the landing of seaplanes, impact problems were enlightened by the concept of pressure impulse introduced by Bagnold \cite{Bagnold1939}. Otherwise this quantity was used by Lamb \cite{LambHydro} in order to develop a mathematical model of suddenly changed flow. The idea is to notice that a sudden change of the motion of one of the boundary of the fluid domain induces pressure gradients which in turn produce a sudden change in the liquid velocity \citep{Batchelor}. This change occurs over a timescale $\tau$ very small compared to the convective time $R/U$. Therefore, by introducing a small parameter $\varepsilon=\frac{\tau}{R/U}$, we deduce after a comparison of the order of magnitude of each terms of the momentum equation that the non-linear terms are negligible compared to the time derivative of the velocity. In this study we only consider inertia-dominated impact then we assume that gravity, capillary and viscous effects are small with respect to inertial ones, \textit{i.e.} Froude $\text{Fr}=U^2/g R$, Weber $\text{We} = \rho U^2 R/\sigma$ and Reynolds $\text{Re} = \rho U R/\mu$ numbers are all large with respect to unity. Here $g$ denotes the gravity, $\sigma$ the liquid-gas surface tension, $\rho$ the liquid density and $\mu$ its viscosity. Consequently the time derivative of the velocity is just balanced by the pressure gradient. Then, at leading-order, the problem is described by the following equation \citep{LambHydro,Batchelor}~:
\begin{gather}
\label{eq:eqg1}
\frac{\partial \boldsymbol{u}}{\partial t} = - \frac{1}{\rho} \boldsymbol{\nabla} p,
\end{gather}
where $p$ is the pressure of the liquid. We assume here that the atmospheric pressure is the reference pressure. By integration of the relation (\ref{eq:eqg1}) on the impact duration, we obtain : 
\begin{gather}
\label{eq:eqg2}
\boldsymbol{u}_{impact}=\boldsymbol{u}(\tau) - \boldsymbol{u}(0) = - \frac{1}{\rho} \boldsymbol{\nabla} P,
\end{gather}
with $P$ the pressure impulse defined by:
\begin{gather*}
P = \int_0^{\tau} p(\boldsymbol{x},t) \text{d} t.
\end{gather*}

The impact velocity considered in this paper is assumed much lower than the speed of sound $c$. Hence we can suppose that the flow induced by the impact is incompressible. Therefore by taking the divergence of (\ref{eq:eqg2}) we deduce that the pressure impulse satisfies Laplace's equation $\Delta P = 0$. 

The problem as described here is general and at this stage the pressure impulse theory could be applied to many problems whatever the geometry \emph{e.g.} with a complete sphere for drop deformation by laser-pulse impact \citep{Gelderblom2016} or with a plane for the impact of a wave on a seawall \citep{Cooker1995}. However the mathematical form of the solution strongly depends on the geometry and on the boundary conditions.

\subsection{Problem statement}
\label{subsec:Pb}
We consider a perfectly hemispherical drop of density $\rho$, surface tension $\sigma$ and radius $R$, lower than the gravity-capillary length $l_{gc}=\sqrt{\sigma/\rho g}$, disposed on a circular cylinder. The base of the hemisphere coincides with the circular disc at the top of the cylinder whose radius is at least $R$. The cylinder impulsively starts from rest with a velocity $U$ toward the drop or equivalently the cylinder falls in free fall and imposes a velocity $U$ to the drop when that one hits the soil. Hence the impact induces a flow assumed axisymmetric and inviscid. As shown in the previous paragraph the impulsive problem is reduced to the derivation of Laplace's equation, in spherical coordinates in the present case: 
\begin{gather}
\frac{1}{r^2} \frac{\partial}{\partial r} \left(r^2 \frac{\partial P}{\partial r} \right) + \frac{1}{r^2 \sin \theta} \frac{\partial}{\partial \theta} \left( \sin \theta \frac{\partial P}{\partial \theta} \right) = 0.
\end{gather}
This equation is completed by \emph{(i)} a dynamical condition which represents normal stress continuity at the free surface $\mathcal{S}$ and which takes account of the high Weber number hypothesis and \emph{(ii)} a condition expressing the sudden variation of velocity $(\boldsymbol{u} \cdot \boldsymbol{e}_z)\vert_{z=0} = U$ at the bottom of the drop $\mathcal{P}$ at the impact moment, where $\boldsymbol{e}_z$ is the upward unit vector. These two conditions are respectively given by the following relations:
\begin{gather}
P(r=R) = 0,  \\
\left.\frac{\partial P}{\partial \theta}\right\vert_{\theta=\frac{\pi}{2}} = \rho U r.
\end{gather}
The problem is non-dimensionalised using the scales $R$, $\rho$ and $U$ and by introducing the following quantities:  
\begin{gather}
r = R \bar{r}, \quad \boldsymbol{u} = U \bar{\boldsymbol{u}}, \quad P = \rho U R \bar{P},
\end{gather}
we rewrite the equations into their dimensionless counterparts:
\begin{gather}
\frac{1}{\bar{r}^2} \frac{\partial}{\partial \bar{r}} \left(\bar{r}^2 \frac{\partial \bar{P}}{\partial \bar{r}} \right) + \frac{1}{\bar{r}^2 \sin \theta} \frac{\partial}{\partial \theta} \left( \sin \theta \frac{\partial \bar{P}}{\partial \theta} \right) = 0, \\
\bar{P}(\bar{r}=1) = 0,  \\
\left.\frac{\partial \bar{P}}{\partial \theta}\right\vert_{\theta=\frac{\pi}{2}} = \bar{r}.
\label{eq:WB_CL}
\end{gather}
\begin{figure}
\begin{center}
\includegraphics[width=9cm]{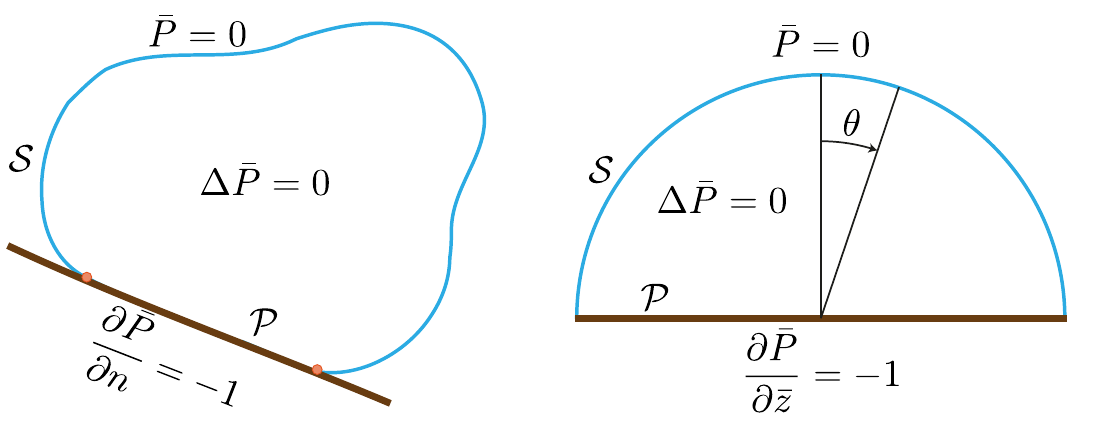}
\caption{Left : Impulsive problem sketched for a liquid domain of arbitrary shape at the impact moment. Right : Representation of the configuration considered in the present paper where the free surface $\mathcal{S}$ is hemispherical.}
\label{fig:ModeleDG}
\end{center}
\end{figure}
\noindent The corresponding model is sketched Fig.~\ref{fig:ModeleDG}. We notice that $\bar{P}_0=-\bar{r} \cos \theta$ is a trivial solution of Laplace's equation which verifies the Neumann boundary condition (\ref{eq:WB_CL}). Then for convenience we introduce a translation of the pressure impulse such that $\bar{P}=-\bar{r} \cos \theta +\tilde{P}$. This translation allows us to represent the solution in the reference frame of the drop. The problem is hence rewritten as the Laplace's equation $\Delta \tilde{P}=0$ with the following boundary conditions:
\begin{gather}
\tilde{P}(\bar{r}=1) = \cos \theta \label{eq:WB_CLTilde_a}
 \\
\left.\frac{\partial \tilde{P}}{\partial \theta}\right\vert_{\theta=\frac{\pi}{2}} = 0.
\label{eq:WB_CLTilde_b}
\end{gather}
This mathematical formulation of the impact problem given by Laplace's equation and mixed boundary conditions is very general and is reminiscent of many other problems involving different geometries \emph{e.g.} the dam break problem \citep{Korobkin2009,Stansby1998}. We will present some other examples for two-dimensional problems in the appendix (section \ref{sec:Appendix}).

\section{Solution of the impulsive problem and determination of the slip velocity}
\label{sec:Resolution}
In spherical coordinates, axisymmetric Laplace's equation can be solved with variable separation, leading to a family of elementary spherical harmonic solutions given by $F_n(\bar{r},\theta) = \bar{r}^n P_n(\cos \theta)$ where $P_n$ are Legendre polynomials of degree $n$. By decomposing the solution into odd and even parts the condition (\ref{eq:WB_CLTilde_b}) imply that the solution of the problem must be written as $\sum_{n=0}^{+ \infty} A_{2n} F_{2n}$ \citep{Antkowiak2007}. Hence the pressure impulse is given by the following relation:
\begin{gather}
\label{s1}
\tilde{P}(\bar{r},\theta) =\sum_{n=0}^{+ \infty} A_{2n} P_{2n}(\cos \theta) \bar{r}^{2n}.
\end{gather}
The coefficients $A_{2n}$ are determined with the Dirichlet boundary condition on the free-surface (\ref{eq:WB_CLTilde_a}) and the orthogonality relations of the Legendre polynomials \citep{Byerly}. Therefore, the pressure impulse in the reference frame of the laboratory or equivalently the velocity potential is given by \citep{Tyvand2014,Philippi2013}~:
\begin{gather}
\label{eqpr}
\bar{P}(\bar{r},\theta) =-\bar{z} + \sum_{n=0}^{\infty} \frac{(-1)^{n+1} (4n+1)}{(2n-1)(2n+2)} \frac{(2n)!}{2^{2n} (n!)^2} P_{2n}(\cos \theta) \bar{r}^{2n}.
\end{gather}
Finally the velocity field in the reference frame of the drop at the impact moment is deduced from relation (\ref{eq:eqg2}):
\begin{gather}
\label{eq:IsoVitRad}
\boldsymbol{u}_{{\text{impact}}_{\vert R_{drop}}}(\bar{r},\theta) = - \boldsymbol{\nabla}
\left(\sum_{n=0}^{\infty} A_{2n} P_{2n}(\cos \theta) \bar{r}^{2n}\right).
\end{gather}

\begin{figure}
\begin{center}
\includegraphics[width=7cm]{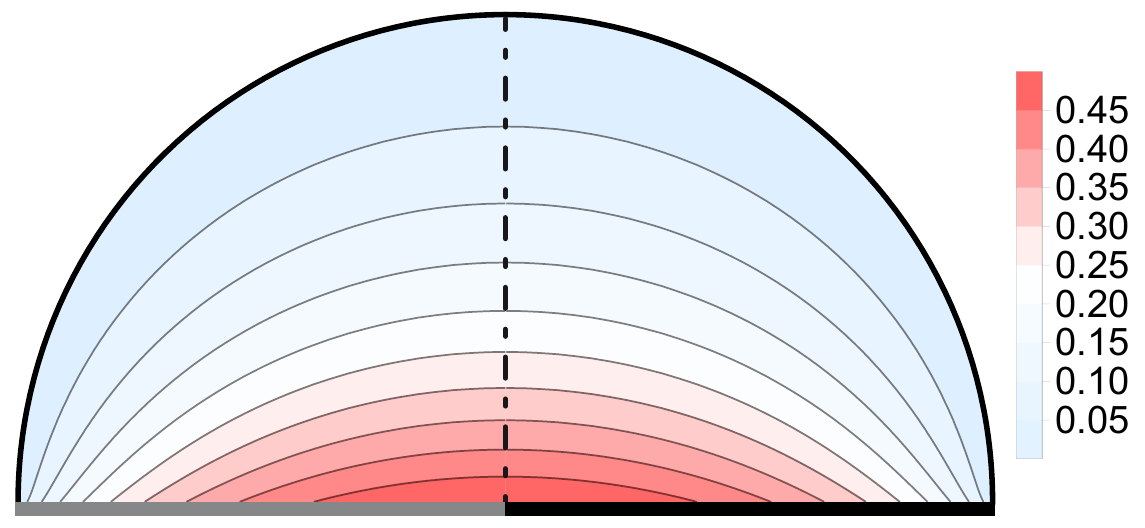}
\caption{Comparison between the pressure impulse field induced by the impact of a liquid hemisphere and a solid substrate extracted from the first step of a \gerris\ simulation (left) and the one represented with equation (\ref{eqpr}) obtained with the pressure impulse theory (right). The pressure field computed with \gerris\ ($\text{Re}=3000$ and $\text{We}=130$) is multiplied by the value of the first time step $\text{dt}=3.333 \times 10^{-5}$ in order to obtain the pressure impulse. As the isovalues are the same for both cases (0.05, 0.1, 0.15, 0.2, 0.25, 0.3, 0.35, 0.4, 0.45, 0.5), we note that theoretical and numerical approaches are in excellent qualitative agreement.}
\label{fig:IsoP}
\end{center}
\end{figure}
\noindent A closed-form expression of these solutions is unfortunately not accessible in the general case. However, we will show in section \ref{subsec:V_P_axe} that it is possible to calculate the value of these series at some particular places.

Fig.~\ref{fig:IsoP} represents a comparison between the structure of the pressure impulse field obtained from equation (\ref{eqpr}) and the one extracted from the first step of a numerical computation performed with the \gerris\ flow solver (freely downloadable at \url{http://gfs.sourceforge.net}). \gerris\ is a solver of the incompressible Navier-Stokes equations for multiphase flows taking into account surface tension and using a finite-volume approach, adaptive mesh refinement to reduce computational costs and a Volume-of-Fluid (VoF) method for interface tracking \citep{Popinet2003,Popinet2009}. The computation was performed in axisymmetric coordinates and the initial configuration corresponds to a liquid hemisphere disposed on a solid surface. The air motion is also computed. The liquid phase is initialised with a constant downward velocity with a slip boundary condition at the substrate level. The Reynolds and Weber numbers corresponding to the impact are respectively of 3000 and 130 in order to perform a simulation in an inertia-dominated regime. As \gerris\ only computes the pressure field it is necessary to multiply this quantity by the value of the first time step $\text{dt}=3.333 \times 10^{-5}$ in order to obtain the pressure impulse field. We verified that the value of the first time step of the simulation does not influence the results. The two approaches are in excellent qualitative agreement and reveal a structure with a maximum pressure at the center of the impact reminiscent of some classic problems involving Laplace's equation with such boundary conditions \textit{e.g} the impact of a wave on a seawall \citep{Cooker1995}.

\begin{figure}
\begin{center}
\includegraphics[width=7cm]{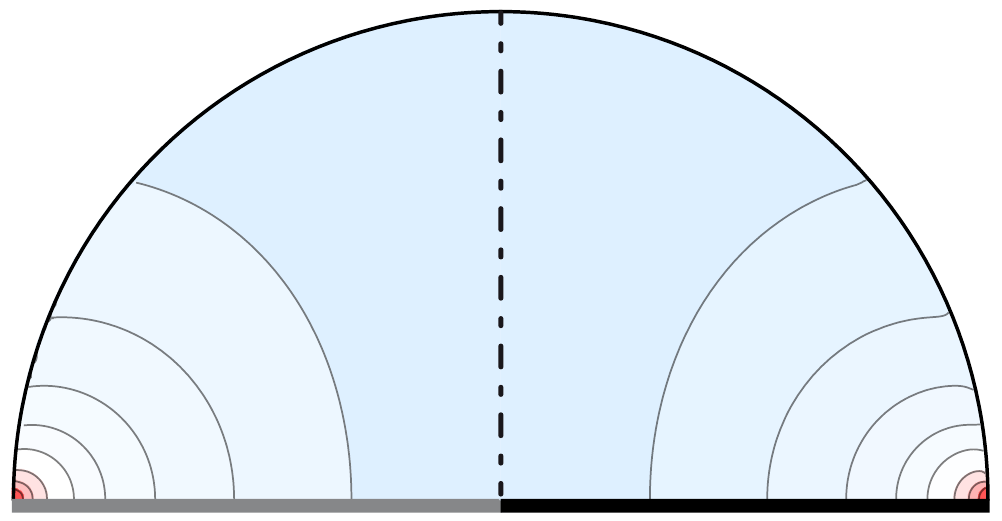}
\caption{Comparison between the radial velocity field induced by the impact of a liquid hemisphere and a solid substrate extracted from the first step of a \gerris\ simulation (left; $\text{Re}=3000$ and $\text{We}=130$) and the one obtained with the theoretical prediction (\ref{eq:IsoVitRad}) deduced from the pressure impulse theory (right). Theoretical and numerical results are in excellent qualitative agreement (isovalues: 0.2, 0.4, 0.6, 0.8, 1, 1.3, 1.6, 2).}
\label{fig:IsoVitRad}
\end{center}
\end{figure}
We represent Fig.~\ref{fig:IsoVitRad} a comparison between the structure of the radial velocity field extracted from numerical computations performed with \gerris\ and the one deduced from the pressure impulse theory via equation (\ref{eq:IsoVitRad}). An excellent qualitative agreement between the theoretical prediction and the numerical simulation is measurable. Interestingly, the structure of this field shows a degenerate behavior near the contact line which is a hint of the existence of a singularity in this region. 

The previous theoretical predictions can be used to determine some quantities of interests such as the total net normal force induced by the impact on the solid substrate or the slip velocity. The first is obtained by integration over the wet surface of the pressure impulse field and will be determined in section \ref{sec:Discussion}. This calculation involves the pressure impulse across the wetted region $\bar{P}(\bar{r}):=\bar{P}(\bar{r},\theta=\frac{\pi}{2})$, directly deduced from equation (\ref{eqpr}):
\begin{gather}
\label{eq:P_r}
\bar{P}(\bar{r}) = - \sum_{n=0}^{\infty} \frac{(4n+1)}{(2n-1)(2n+2)} \left(\frac{(2n)!}{2^{2n} (n!)^2} \right)^2 \bar{r}^{2n}.
\end{gather}
\begin{figure}
\begin{center}
\includegraphics[width=7cm]{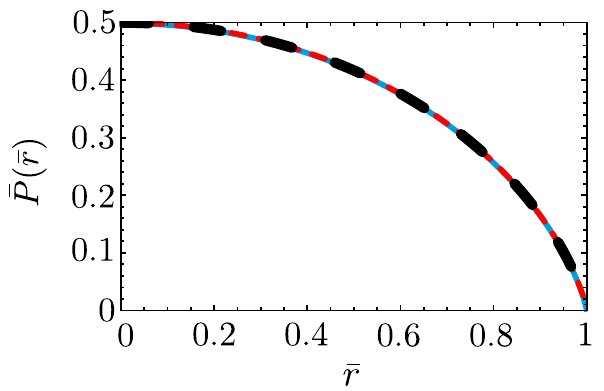}
\caption{Representation of the analytic solution (red dashed line) of the pressure impulse along the wet radius in excellent agreement with numerical solutions extracted from \gerris\ (blue solid line) and \ff\ (black dashed line). Just as Fig.~\ref{fig:IsoP} we obtained the pressure impulse by multiplying the pressure field computed with \gerris\ ($\text{Re}=3000$ and $\text{We}=130$) by the value of the first time step $\text{dt}=3.333 \times 10^{-5}$.}
\label{fig:PressionIntFront}
\end{center}
\end{figure}
A representation of this last solution in very good agreement with numerical solutions obtained with \gerris\ and \ff\ \citep{Hecht2012} is represented Fig.~\ref{fig:PressionIntFront}. This last software (freely downloadable at \url{http://www.freefem.org}) is an open-source code solving partial differential equations using the finite element method. The computation was performed with the same initial configuration than the \gerris\ one and the Laplace's equation with the appropriate boundary conditions was directly solved. The relative error between the \gerris\ numerical solution and the theoretical prediction is lower than 1 \% except close to the contact line where the error increases up to 6 \%. On the other hand the relative error is lower than 1 \% all along the wetted region in the case of the \ff\ numerical solution. As suggested by Fig.~\ref{fig:IsoP}, the maximum of the pressure impulse is located on the axis of symmetry of the drop leading to a structure of the pressure field reminiscent of the one generated by a wave impacting a seawall \citep{Cooker1995}. This kind of structure is not universal for impact problems and the pressure maximum could be located close to the contact line just like \emph{e.g.} in the case of the water entry problem \citep{Wagner1932,Cointe1989,Howison1991} or for drop impact \citep{Philippi2016}.

The slip velocity or velocity at the edge $\bar{u}_e(\bar{r}):=\bar{u}_r(\bar{r},\theta=\frac{\pi}{2})$ is an other quantity of interest. It is directly deduced from equation (\ref{eq:IsoVitRad})~:
\begin{eqnarray}
\label{eq:vr}
\bar{u}_e(\bar{r}) & = & \sum_{n=1}^{\infty} \frac{n(4n+1)}{(2n-1)(n+1)} \left(\frac{(2n)!}{2^{2n} (n!)^2}\right)^2 \bar{r}^{2n-1}.
\end{eqnarray}
\begin{figure}
\begin{center}
\includegraphics[width=7cm]{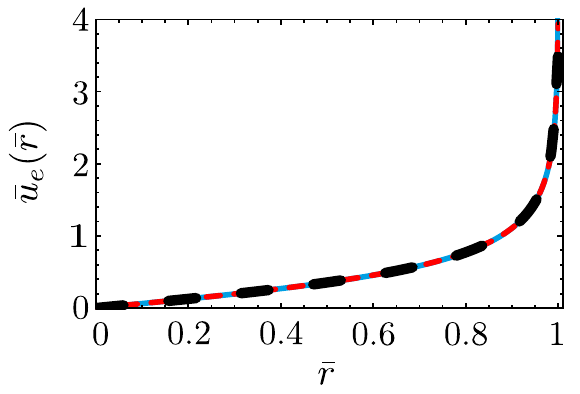}
\caption{Comparison of the theoretical prediction of the slip velocity (red dashed line) with numerical solutions extracted from \gerris\ (blue solid line - $\text{Re}=3000$ and $\text{We}=130$) and \ff\ (black dashed line).}
\label{fig:vrIntFront}
\end{center}
\end{figure}
Although this inviscid solution is not physical it is nonetheless relevant for it corresponds to the edge velocity of the viscous boundary layer. More precisely this quantity is involved in the composite solution matching the inviscid and viscous solutions \cite{VanDyke1975}. This theoretical prediction is compared with numerical solutions computed with \gerris\ and \ff\ in Fig.~\ref{fig:vrIntFront}, revealing an excellent agreement between the different approaches. The relative error between both \gerris\ and \ff\ numerical solutions and the theoretical prediction is around 1 \%. The nature of the solution (\ref{eq:vr}) and the structure of the velocity field will be discussed in section \ref{sec:Discussion}. Nonetheless the boundary layer problem is beyond the scope of the present paper. We note that solutions (\ref{eq:P_r}) and (\ref{eq:vr}) can be expressed with hypergeometric functions that we will not specify here.

\section{Discussion}
\label{sec:Discussion}
\subsection{Singularity at the contact line of the slip velocity}
\label{subsec:singularite}
As already mentioned in the previous section, the radial velocity field has a singular behavior close to the contact line. In order to discover the nature of this singularity we will determine the asymptotic behavior of the slip velocity deduced from spherical harmonics near the contact line (see equation (\ref{eq:vr})).

By using Stirling's approximation we deduce that the general term of this series is asymptotically equivalent to $\frac{2}{\pi n}$. Hence close to the contact line, the slip velocity (\ref{eq:vr}) is given by the relation:
\begin{gather*}
\bar{u}_e(\bar{r}) \underset{\bar{r} \to 1}{\sim} \frac{2}{\pi} \sum_{n=1}^{\infty} \frac{\bar{r}^{2n-1}}{n}.
\end{gather*}
Recognizing the Taylor expansion of the logarithm we finally deduce that the asymptotic behavior of the slip velocity close to the contact line corresponds to a logarithmic divergence leading to the following approximation:
\begin{gather}
\label{equiv}
\bar{u}_e(\bar{r}) \underset{\bar{r} \to 1}{\sim} -\frac{2}{\pi} \frac{\log(1-\bar{r}^2)}{\bar{r}}.
\end{gather}
\begin{figure}
\begin{center}
\includegraphics[width=7cm]{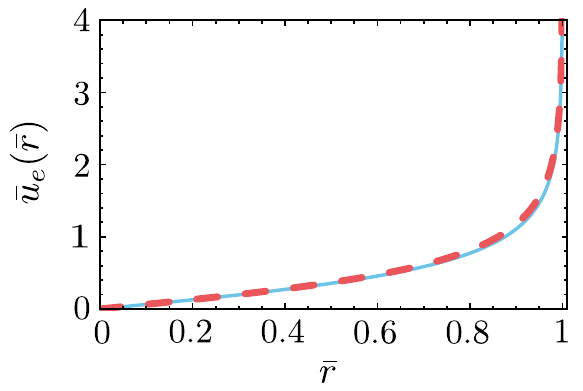}
\caption{Comparison between the analytical solution of the slip velocity deduced with spherical harmonics (equation (\ref{eq:vr}) - blue continuous line) and its logarithmic equivalent close to the contact line (equation (\ref{equiv}) - red dashed line).}
\label{fig:vitesse_log}
\end{center}
\end{figure}
This last asymptotic equivalent matches satisfactorily with the theoretical prediction even near the origin (see Fig.~\ref{fig:vitesse_log}). We note that each of the numerical solutions deviate from the singular theoretical prediction near the contact line and reach a finite value. The logarithmic singularity of the slip velocity is in fact not really surprising. Mathematically, a singularity at the contact line of the gradient of a harmonic function is a consequence of the transition from Dirichlet to Neumann boundary condition at this location. This kind of behavior is universal in potential theory and is reminiscent of numerous problems as heat transfer, fracture mechanics with the thorn singularity \citep{Leguillon} or fluid mechanics \citep{Stansby1998,Rayleigh1911,Penney1952}. Depending on the shape of the domain, the nature of the singularity could also evolve into a power-law singularity as for the problem of droplet's evaporation \citep{Deegan1997}.

The regularisation of the singularity will be done in a future work. However, it is clear that it is not possible to neglect the non-linear effects close to the contact line because of the high velocity gradient leading to formation of a liquid sheet in this region. In such a case, equation (\ref{eq:eqg1}) would be not valid anymore and the pressure impulse theory would be not accurate in this region. Consequently, a more careful investigation is necessary to resolve this singularity. Indeed, the solution presented in this paper is an outer solution. The correct structure of the flow field near the contact line should be revealed by using the method of matched asymptotic expansions. More precisely it will be relevant to follow the methodology from \emph{e.g} Korobkin and Yilmaz \cite{Korobkin2009}, King and Needham \cite{King1994} or Needham \emph{et al.} \cite{Needham2007}. The idea is to define at least one appropriate inner region and consider the correction of the leading-order problem. The new problem will be nonlinear and its solution will be spatially non-singular.

\subsection{Closed-form expressions for the pressure impulse and velocity fields along the axis of the drop}
\label{subsec:V_P_axe}
We represented in the previous section the solutions of the impact problem with spherical harmonics. A closed-form expression is unfortunately not accessible in the general case. However, there exists explicit solutions for the pressure and velocity fields along the axis of symmetry of the drop. These two quantities, depending on the coordinates $\bar{z} = \bar{r}$ along this axis, are deduced from equations (\ref{eqpr}) and (\ref{eq:IsoVitRad}) for $\theta=0$ and are respectively given by:
\begin{eqnarray}
\bar{P}(\bar{z}):=\bar{P}(\bar{z},\theta=0) = -\bar{z} + \sqrt{1+\bar{z}^2} + \frac{1-\sqrt{1+\bar{z}^2}}{\bar{z}^2}, \label{eq:Axe1}\\
\bar{u}_z(\bar{z}):=\bar{u}_z(\bar{z},\theta=0) = \frac{-2-\bar{z}^2-\bar{z}^4 + 2 \sqrt{1+\bar{z}^2}}{\bar{z}^3 \sqrt{1+\bar{z}^2}} \label{eq:Axe2}
\end{eqnarray}
\begin{figure}
\begin{center}
\includegraphics[width=9cm]{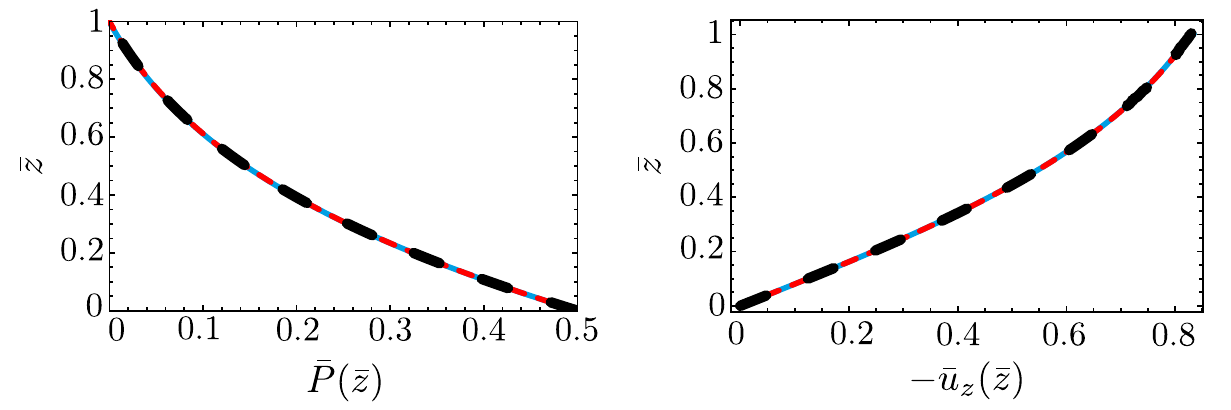}
\caption{Representation of the analytic solution (red dashed line) of respectively the pressure impulse (left) and the velocity (right) along the axis of the drop in excellent agreement with numerical solutions extracted from Gerris (blue solid line) and FreeFem++ (black dashed line). The pressure impulse was obtained by multiplying the pressure field computed with Gerris (Re = 3000 and We = 130) by the value of the first time step $\text{dt}=3.333 \times 10^{-5}$. }
\label{fig:P_V_Axe}
\end{center}
\end{figure}
These last results are confronted with numerical solutions extracted from \gerris\ and \ff\ in figure \ref{fig:P_V_Axe}. There is again an excellent agreement between the computations and the theoretical prediction.

\subsection{Structure of stagnation point flow}
\begin{figure}
\begin{center}
\includegraphics[width=7cm]{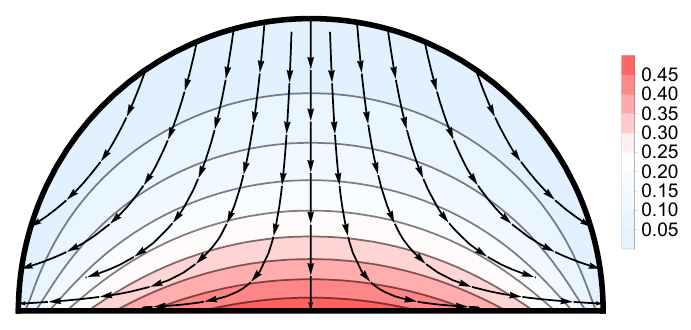}
\caption{Streamlines of the flow induced by the impact extracted from equation (\ref{eq:IsoVitRad}). The pressure field developing inside the drop is represented in the background.}
\label{fig:PA_DG}
\end{center}
\end{figure}

The calculation of the first mode of the velocity field (\ref{eq:IsoVitRad}) respectively along the substrate ($\theta = \pi/2$) and along the axis of
symmetry of the drop ($\theta=0$) leads to the following representation of the flow near the origin:
\begin{gather}
\label{hiemenz}
\left\lbrace \begin{array}{ccc}
\bar{u}_r(\bar{r},\bar{z}) & \simeq & \frac{5}{8} \bar{r},\\
\bar{u}_z(\bar{r},\bar{z}) & \simeq & -\frac{5}{4} \bar{z},\\
\end{array} \right.
\end{gather}
which corresponds to a structure of stagnation point flow. Equivalently, the $\bar{u}_z$ component of the field could be obtained with the first order power series of the analytic solution (\ref{eq:Axe2}). The complete structure of the flow within the drop at the impact moment is extracted from equation (\ref{eq:IsoVitRad}) and represented Fig.~\ref{fig:PA_DG} via its streamlines. This last reveals the stagnation point structure of the flow.

By applying the same series expansion to the asymptotic equivalent of the slip velocity (\ref{equiv}) near the origin, we obtain the following relation:
\begin{gather*}
-\frac{2}{\pi} \frac{\log (1- \bar{r}^2)}{\bar{r}} = \frac{2}{\pi} \bar{r} + o(\bar{r}^2), \quad \bar{r} \rightarrow 0.
\end{gather*}
Consequently the good agreement between equation (\ref{equiv}) and the slip velocity near the origin of the impact is due to the well known rough approximation $\frac{5}{8} \simeq \frac{2}{\pi}$. 

\subsection{Momentum's lost during impact}
By applying Newton's second law to the drop, we obtain by integration over the impact duration $\tau$ the momentum's variation at the impact moment :
\begin{gather}
\label{eq:qdm}
\bar{\boldsymbol{p}}_1 - \bar{\boldsymbol{p}}_0 = \int_0^{\tau} \bar{\boldsymbol{F}}(t) \,\text{d}t,    
\end{gather}
where $\bar{\boldsymbol{p}}_0$ and $\bar{\boldsymbol{p}}_1$ are respectively the momentum of the drop before and after the impact. The net force $\bar{\boldsymbol{F}}$ applied to the system corresponds here to the net normal total force exerted by the boundary on the fluid. The latter is given by the integration of the pressure field over the wet surface. We deduce after a permutation of spatial and temporal integration that the vertical momentum's variation directly depends on the integration of pressure impulse over the wet length $\mathcal{P}$. Hence the projection of the equation (\ref{eq:qdm}) onto $e_z$ is given by the following scalar equation:
\begin{gather*}
\bar{p}_0 - \bar{p}_1  = \int_{\mathcal{P}} \bar{P}(\bar{r})\,\text{d}S,
\end{gather*}
with $\text{d}S=\bar{r} \, \text{d} \bar{r} \, \text{d} \varphi$. Therefore equation (\ref{eq:P_r}) imply that the difference of vertical momentum is given by :
\begin{eqnarray*}
\bar{p}_0 - \bar{p}_1 &=& - 2 \pi \sum_{n=0}^{+ \infty} \frac{(4n+1)}{(2n-1)(2n+2)^2} \left(\frac{(2n)!}{2^{2n} (n!)^2}\right)^2\\
 &=& \frac{2}{3} (3 \pi-8).
\end{eqnarray*}
On the other hand, the initial momentum is simply $\bar{p}_0=\frac{2}{3} \pi$. Finally the resulting momentum after impact in its dimensional counterpart is given by :
\begin{gather}
p_1=\int_{\mathcal{V}} \rho \boldsymbol{u}.\boldsymbol{e}_z \,\text{d} V = \frac{2}{3} (8-2\pi) \,\rho U R^3.
\end{gather}

\section{Conclusion}
We considered in this paper impact problems involving hemispherical liquid domains, with an emphasis on its impulsive aspect. The initial configuration we consider in this study is the free fall of a rigid cylinder with a drop disposed at its top. A vertical velocity is imposed to the drop when this last hits the soil. 

By using the pressure impulse theory this problem could be reduced to the derivation of Laplace's equation with Neumann and Dirichlet boundary conditions. The pressure impulse along the wetted region and the slip velocity have been deduced from the spherical harmonics representation of the solution. The theoretical predictions proposed in this paper are in very good agreement both qualitatively and quantitatively with the first time step of \gerris\ numerical simulation and with \ff\ calculation.

The structure of the velocity field induced by the impact has been depicted in the discussion. In particular we exhibit a logarithmic singularity at the contact line for the slip velocity, due to the mixed boundary conditions at this location. We also determined closed-form expressions for the pressure impulse and velocity fields along the axis of the drop and we discussed the structure of stagnation point of the flow. To complete the description of the flow induced by the impact it will be necessary to match the inviscid solution with the viscous one and to explain the regularisation of the velocity field close to the contact line. 

The impact problem as defined in section~\ref{subsec:Pb} is general and could be used to study many configurations with different geometries as detailed in the appendix (section~\ref{sec:Appendix}) for two-dimensional problems involving planar and circular liquid domains. The structure of the pressure impulse and velocity fields are analogous to those determined in the axisymmetric case.

\appendix

\section*{Appendix : The two-dimensional case}
\label{sec:Appendix}
As mentioned in section \ref{sec:Model}, the class of impact problems studied in this paper is very general. In particular the pressure impulse theory could be applied in many problems involving different geometries of the fluid domain. In this appendix we will solve few two-dimensional impulsive problems involving plane and circular boundaries. Since the problem is reduced to the derivation of Laplace's equation, several methods could be used, leading to equivalent mathematical representations of the solution of the problem (see section \ref{sec:Resolution}). Complex analysis is especially an interesting alternative for two-dimensional problems.

\section{Short-time behavior of free-surface flows generated by moving a vertical plate}
\begin{figure}
\begin{center}
\includegraphics[width=9cm]{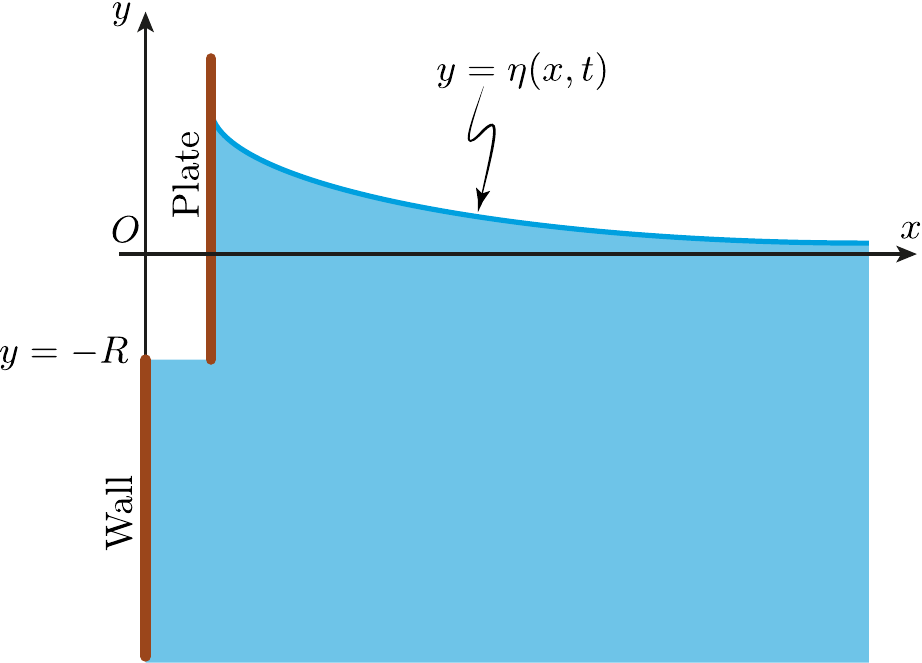}
\caption{Sketch of the problem studied by Roberts \citep{Roberts1987} where the horizontal impulsive motion of a vertical plate generates free-surface flow.}
\label{fig:rob}
\end{center}
\end{figure}
We consider here the following problem studied by Roberts \citep{Roberts1987}. An infinitely deep fluid with a free surface $y=\eta(x,t)$ is bounded on the left by a vertical wall at rest. As shown Fig.~\ref{fig:rob}, the plate of height $R$ initially moves horizontally with a constant velocity, corresponding to an impulsive acceleration as described in the original paper. The flow induced by this motion being irrotational we can define a velocity potential $\phi$ such as the velocity field is given by $\boldsymbol{u} = \nabla \phi$. By assuming that the flow is incompressible we deduce that the velocity potential is solution of Laplace's equation $\Delta \phi = 0$. An approach to determine the evolution of the flow near the contact line is to use a time decomposition in series of time for $\phi$ and $\eta$ \citep{Peregrine1972} :
\begin{gather*}
\phi = \phi_0 + t \, \phi_1 + \cdots, \quad \eta = \eta_0 + t \, \eta_1 + \cdots.
\end{gather*}
Then the short-time behavior is given by the dominant problem and the velocity potential don't have any time dependence at leading order. We deduce from the analysis of Peregrine \citep{Peregrine1972} that the fluid's motion is governed by Laplace's equation for the velocity potential at leading order $\Delta \phi_0 =0$ in the fluid domain. The uniform velocity of the moving plate leads to the following Neumann boundary condition:
\begin{gather}
\label{eqflu4}
\left. \frac{\partial \phi_0}{\partial x}\right\vert_{x=0} = \left\lbrace \begin{array}{ll}
\alpha, & -R<y<0,\\
0, & y<-R.\\
\end{array} \right.
\end{gather}
where $\alpha$ is a real number purely positive with the dimension of a velocity. At the free surface, Bernoulli's equation implies that the pressure is constant at leading order. Hence we have:
\begin{gather}
\label{eqflu6}
\phi_0 = 0, \quad \text{when} \ y = 0.
\end{gather}
The Peregrine's decomposition is also applied to the kinematic condition at the free-surface $\frac{DS}{Dt} = 0$ with $S = y - \eta(x,t)$. Then at leading-order the free-surface elevation is given by the derivative of the velocity potential:
\begin{gather}
\label{eqflu5}
\eta_1 = \frac{\partial \phi_0}{\partial y}, \quad \text{when} \ y = 0.
\end{gather}
Finally we have to solve Laplace's equation with Neumann and Dirichlet boundary conditions (\ref{eqflu4}) and (\ref{eqflu6}). This problem is formally identical to the one we solved in the present paper where the velocity potential and the free-surface elevation are respectively the counterpart of the pressure impulse and the radial velocity along the wet surface. We also note that the pressure impulse $P$, associated to the impact problem previously defined, may be related to the harmonic velocity potential via the relation $P=-\rho \phi$ \citep{Batchelor}. Then the relation (\ref{eq:eqg2}), which relates the impact velocity with the pressure impulse, corresponds to the relation (\ref{eqflu4}). Therefore we expect that the solutions of the problem described by Roberts will share the same characteristic than the impact one \emph{e.g.} the singular behavior of the derivative of the harmonic function, corresponding to the free-surface elevation in the present case, near the contact line.

The logarithmic behavior of the displacement of the free-surface is determined by Roberts with complex analysis. The idea is to calculate the complex velocity 
by distributing sources along the wall and images above the line $y=0$: \[\bar{q}_0 = \int_{-R}^0 \left(\frac{D}{2 \pi (z-ia)} - \frac{D}{2 \pi (z+ia)} \right)
\text{d}a,\]
where $z=x+iy$ and the source strength on the wall is $D = -2 \left. \frac{\partial \phi_0}{\partial x}\right\vert_{x=0}$. Moreover the complex velocity could be written $\bar{q}_0 = u_0 -i v_0$ by definition. Hence the leading-order short-time free-surface elevation is deduced by identification:
\begin{gather}
\eta_1(x,0)= \frac{\alpha}{\pi} \log\left(1+\frac{R^2}{x^2}\right).
\label{eq:eta0}
\end{gather}
As expected this quantity has a logarithmic behavior near the contact line. This singularity has been removed by Needham \emph{et al.} \cite{Needham2007} by using matched asymptotic expansions.

In the case, not considered by Roberts, where we have to solve this problem with an additional boundary in $y=-R$ with a Neumann condition $\frac{\partial \phi}{\partial y}=0$, we need to symmetrize by this new line the solution obtained with complex analysis. The idea is to use an infinity of images along the $x$-axis. We calculate the complex velocity with the image by the line $y=-R$ of the distribution of source along the wall and its image. And by induction, we calculate the complex velocity with the image by the line $y=-n R$ and $y=n R$ where $n$ is an odd number and we sum all of these contributions: 
\begin{gather*}
\bar{q}_0 = \frac{\alpha}{\pi} \int_{0}^R \left(\frac{1}{z+ia} - \frac{1}{z-ia} + \frac{1}{z+2iR-ia} - \frac{1}{z+2iR+ia} +...\right) \text{d}a
\end{gather*}
Then we determine the position of the free-surface by identification: 
\begin{gather}
\label{eq:RobertsSum}
\eta_1(x,0)= \frac{\alpha}{\pi} \sum_{k=-\infty}^{+\infty} (-1)^k \,\text{Im} \left(i \log \left(1+ \frac{R^2}{(x+2 i k R)^2} \right) \right).
\end{gather}
Finally this example illustrates the interest of complex analysis for two-dimensional problems. However, this framework is not applicable in three dimension. From now we will solve problems with methods involving only real analysis, as in the present paper.

\section{Impact of a wave on a seawall : an other example of planar geometry}
\begin{figure}
\begin{center}
\includegraphics[width=9cm]{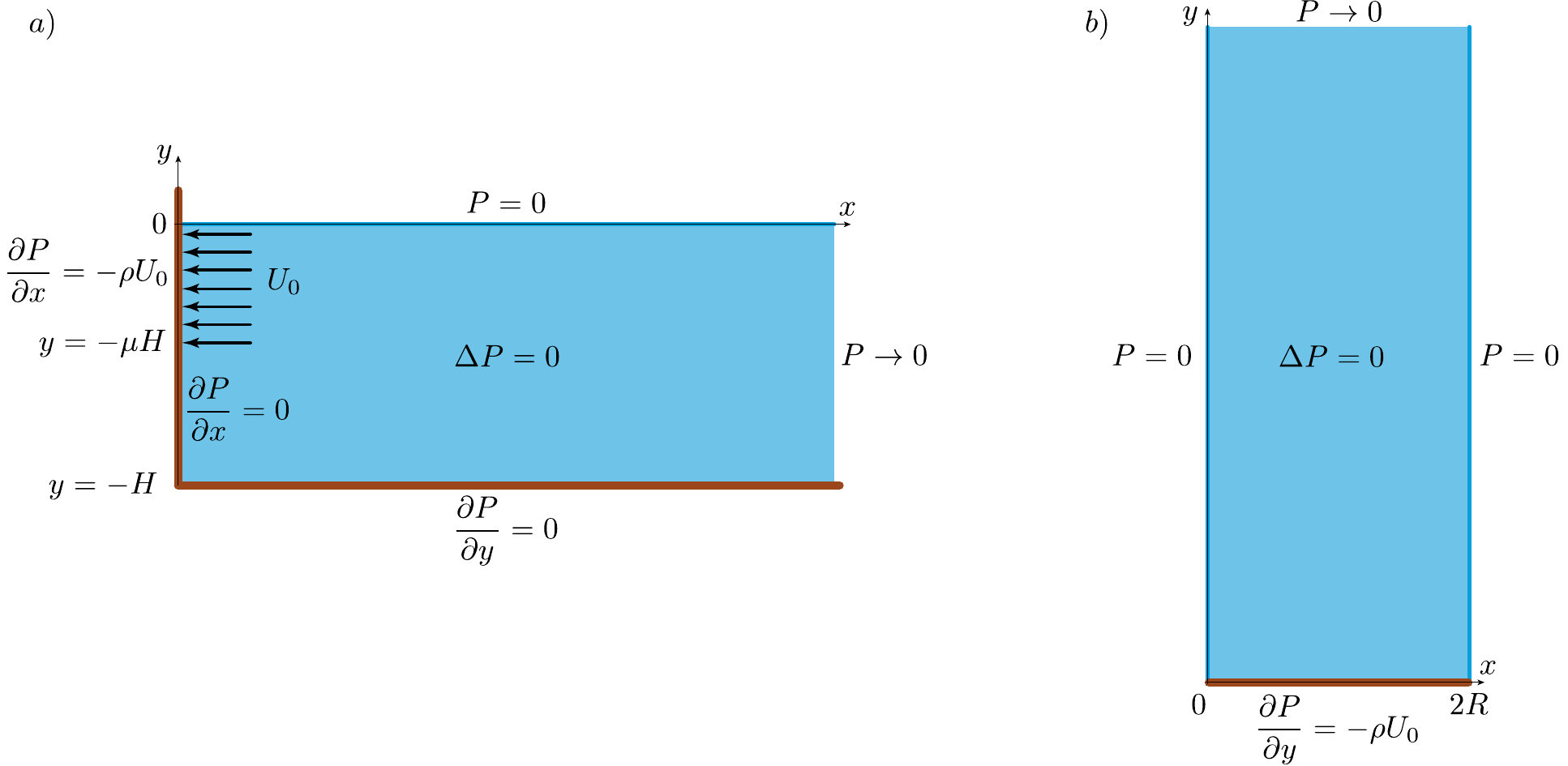}
\caption{a) : Impact of a semi-infinite wave on a fraction $\mu$ of the height of a seawall studied by Cooker and Peregrine \cite{Cooker1995}. A condition of symmetry given by a Neumann boundary condition $\partial P/\partial y=0$ is imposed at the bottom of the fluid domain. b) : Model of the impact of a semi-infinite wave on a solid substrate analogous to the problem solved in the present paper. In this case we have the Dirichlet boundary condition $P=0$ on the right side.}
\label{fig:f4}
\end{center}
\end{figure}

The pressure impulse theory was also applied to the classic problem of the impact of a wave on a vertical seawall by Cooker and Peregrine \cite{Cooker1995}. Just as in the present study, the problem is reduced to the derivation of Laplace's equation with mixed boundary conditions. An extra Neumann condition for the pressure impulse is imposed at the bottom of the domain for symmetry. In the case of an idealized semi-infinite wave impacting a wall on a fraction $\mu$ of its total height (see Fig.~\ref{fig:f4} left) the authors determined with Fourier analysis an analytical solution for the velocity along the wall after impact $v_{impact}$. 
This last quantity, determined with the gradient of the pressure impulse (see equation (\ref{eq:eqg2})), is given in absolute value by: \begin{gather*}
\vert v_{impact}(x,y) \vert =  2 U \sum_{n=1}^{\infty} \frac{\cos(\mu \lambda_n) -1}{\lambda_n} \cos \left(\frac{\lambda_n y}{R} \right) 
\mathrm{e}^{\frac{-\lambda_n x}{R}},
\end{gather*}
with $\lambda_n =\left(n-\frac{1}{2}\right) \pi$. Note that there is a typo in the paper of Cooker and Peregrine \cite{Cooker1995}, a factor -2 is lacking in the equation (4.1). By using the substitutions $x=R \bar{x}$, $y=R \bar{y}$ and $v_{impact}=U \bar{v}_{impact}$, we obtain the dimensionless form of the previous result:
\begin{gather}
\label{vitCP}
\vert \bar{v}_{impact}(\bar{x},\bar{y}) \vert = 2 \sum_{n=1}^{\infty} \frac{\cos(\mu \lambda_n) -1}{\lambda_n} \cos \left(\lambda_n \bar{y}
\right) \mathrm{e}^{-\lambda_n \bar{x}}.
\end{gather}
By replacing the Neumann boundary condition $\partial P/\partial y = 0$ at the bottom of the fluid domain by a Dirichlet condition $P=0$ and choosing $\mu = 1$ we model an impact of a semi-infinite wave on a solid substrate which could be seen as a two-dimensional counterpart for planar geometry of the problem we considered in this paper. By switching axis, as shown Fig.~\ref{fig:f4} right, we deduced from the previous solution and with an argument of symmetry that for all $(\bar{x},\bar{y}) \in \Omega_1 \setminus (0,0)$ (respectively $(\bar{x},\bar{y}) \in \Omega_2 \setminus(2,0)$) the horizontal impact velocity is given by:
\begin{subnumcases}{\bar{v}_{impact}(\bar{x},\bar{y}) =}
-2 \sum_{n=1}^{\infty} \frac{\mathrm{e}^{-\lambda_n \bar{y}}}{\lambda_n} \cos \left(\lambda_n \bar{x} \right), \label{vit_squaredrop1}\\ 
2 \sum_{n=1}^{\infty} \frac{\mathrm{e}^{-\lambda_n \bar{y}}}{\lambda_n} \cos \left(\lambda_n (2-\bar{x}) \right).\label{vit_squaredrop2}
\end{subnumcases}
with $\Omega_1 = [0,1] \times \mathbb{R}_+$ and $\Omega_2 = [1,2] \times \mathbb{R}_+$. The pressure field inside the fluid domain could be also expressed in a similar way for all $(\bar{x},\bar{y}) \in \Omega_1$ (respectively $(\bar{x},\bar{y}) \in \Omega_2$) :
\begin{subnumcases}{\bar{P}(\bar{x},\bar{y}) =}
2 \sum_{n=1}^{\infty} \frac{\mathrm{e}^{-\lambda_n \bar{y}}}{\lambda_n^2} \sin \left(\lambda_n \bar{x} \right),\label{pr_squaredrop1} \\ 
2 \sum_{n=1}^{\infty} \frac{\mathrm{e}^{-\lambda_n \bar{y}}}{\lambda_n^2} \sin \left(\lambda_n (2-\bar{x}) \right).\label{pr_squaredrop2}
\end{subnumcases}
\begin{figure}
\begin{center}
\includegraphics[width=9cm]{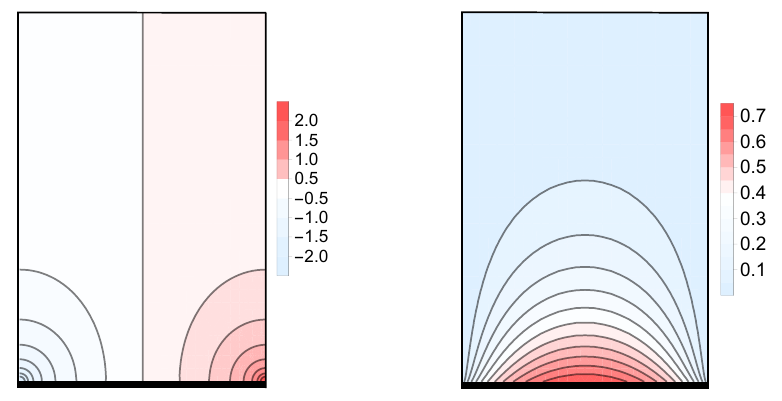}
\caption{Left : Isovalues of the horizontal velocity field (left) and of the pressure impulse field (right) induced by the impact of a semi-infinite wave on a seawall, obtained respectively with the theoretical solutions (\ref{vit_squaredrop1}-\ref{vit_squaredrop2}) and (\ref{pr_squaredrop1}-\ref{pr_squaredrop2}).}
\label{fig:gttecarree}
\end{center}
\end{figure}
We represent Fig.~\ref{fig:gttecarree} the pressure and the radial velocity fields obtained with previous analytical solutions. The height of the domain is large enough to avoid confinement effects in order to be consistent with the hypothesis of semi-infinity. As expected, the structure of these fields are similar than the ones obtained in the case of a hemispherical liquid domain. The maximum value of the pressure impulse occurs at the center of the wet surface while the horizontal velocity is singular in the corner due to mixed boundary conditions at this point as explained in the section \ref{subsec:singularite}.

Cooker and Peregrine \cite{Cooker1995} show in their appendix by using complex analysis that the singularity near the wall is logarithmic ($\bar{v}_{impact}(0,\bar{y}) \sim - \frac{2}{\pi} \log(\bar{y})$). Up to this point, we have completely followed the analysis of Cooker and Peregrine \cite{Cooker1995}. However, contrary to these authors we want to determine the nature of the singularity at the corner with a method involving real analysis. We consider from now the solution corresponding to $1 \leq \bar{x} \leq 2$. In particular we study the solution along the free surface $\bar{x}=2$. Hence the horizontal component of the velocity field is given by:
\begin{gather}
\label{vimpact}
\bar{v}_{impact}(2,\bar{y})= 2 \sum_{n=1}^{\infty} \frac{\mathrm{e}^{-\lambda_n \bar{y}}}{\lambda_n}.
\end{gather}
In order to show that this sum have a logarithmic singularity, we first consider the following equation, valid for $y>0$ :
\begin{gather*}
\sum_{n=1}^{+ \infty} 2 \mathrm{e}^{-\lambda_n y} = \text{cosech}\left(\frac{\pi y}{2}\right).
\end{gather*}
By integrating this relation from $y=\bar{y}$ to $y=+\infty$, we obtain :
\begin{gather}
\bar{v}_{\text{impact}}(2,\bar{y}) = - \frac{2}{\pi} \log \left[\text{tanh} \left(\frac{\pi \bar{y}}{4}\right) \right].
\end{gather}
Finally, the velocity along the free-surface has a logarithmic singularity as $\bar{y}$ tends to zero.

\section{Derivation of the two-dimensional hemispherical liquid impact problem}

In this last appendix we propose to solve the impact problem studied in this paper in a two-dimensional space by using boundary integral method. This problem is also analogous to the one studied by Cooker and Peregrine \cite{Cooker1995} for a circular liquid domain. The boundary of the domain is designated in this case by $C = C_1 \cup C_2$ with $C_1=\lbrace (x,y)\vert -R \leq x \leq R, y=0\rbrace$ and $C_2=\lbrace (x,y)\vert x^2+y^2=R^2, y>0 \rbrace$. As the axisymmetric case the model is sketched figure~\ref{fig:ModeleDG} right. 

\subsection{The boundary integral method}
The boundary integral method is used to solve linear partial differential equations and allows to obtain analytical solutions when the geometry is simple \citep{Pozrikidis1997,Pozrikidis2002,Bonnet}. When the shape of the boundaries becomes too complex, numerical solutions can also be computed. The main idea of this method lies in the determination of the value of a field satisfying a certain linear partial differential equation in all points of a domain $\Omega$ only from its values on the boundary. This approach is particularly interesting to solve Laplace's equation. Formally, considering $G$ a Green's function \emph{i.e} such as $\Delta G=\delta$ where $\delta$ is the Dirac delta function and using Green's theorem, the value of a field $f$ for all $\boldsymbol{x}=(x,y,z) \in \Omega^3$ is given by \citep{Pozrikidis1997,Pozrikidis2002,Bonnet}:
\begin{gather}
\label{eqint2}
f(\boldsymbol{x}) = - \int_{\partial \Omega} G( \boldsymbol{x},\boldsymbol{\xi}) \frac{\partial f}{\partial n}(\boldsymbol{\xi})\,\text{d} \xi 
\,\text{d} \zeta  + \int_{\partial \Omega} f(\boldsymbol{\xi}) \frac{\partial G}{\partial n}(\boldsymbol{x},\boldsymbol{\xi}) \,\text{d} \xi 
\,\text{d} \zeta,
\end{gather}
where $\boldsymbol{\xi} =(\xi,\zeta,\tau) \in \partial \Omega^3$.  The unit normal $\boldsymbol{n}$ is, by convention, pointing outward $\Omega$. The first integral of the right hand side of this equation is designated by SLP (single layer potential) and the second integral by DLP (double layer potential). When $\boldsymbol{x} \in \partial \Omega$ we can also compute $f$. However, in this case DLP becomes improper but this term is still convergent \citep{Pozrikidis1997,Pozrikidis2002}. Then for all $\boldsymbol{x} \in \partial \Omega$:
\begin{gather}
\label{eq:IntFront}
f(\boldsymbol{x}) = - 2 \int_{\partial \Omega} G(\boldsymbol{x},\boldsymbol{\xi}) \frac{\partial f}{\partial n}(\boldsymbol{\xi}) \,\text{d} \xi 
\,\text{d} \zeta + 2 \int_{\partial \Omega} f(\boldsymbol{\xi}) \frac{\partial G}{\partial n}(\boldsymbol{x},\boldsymbol{\xi}) \,\text{d} \xi 
\,\text{d} \zeta.
\end{gather}

\subsection{Application to circular liquid impact problem}
\label{subsec:DefGreen}
In order to compute the pressure impulse we have to choose an appropriate Green's function $G$ for this problem. By appropriate we mean that $G$ have to respect some conditions of symmetry. We define two kinds of Green's functions for a given $\boldsymbol{x} \in \bar{\Omega}$ :
\begin{enumerate}
\item the Green's function of first kind, defined by :
\begin{gather*}
\forall \boldsymbol{\xi} \in C, \quad G(\boldsymbol{x},\boldsymbol{\xi}) = 0,
\end{gather*}
\item the Green's function of second kind, defined by :
\begin{gather*}
\forall \boldsymbol{\xi} \in C, \quad \frac{\partial G}{\partial n} (\boldsymbol{x},\boldsymbol{\xi}) = 0,
\end{gather*}
\end{enumerate}
where $C = \mathcal{S} \cup \mathcal{P}$ is the boundary of the domain (see Fig.~\ref{fig:ModeleDG}). We have to choose for all part of the domain a Green's function of first kind or second kind. Because $\bar{P} = 0$ on the free surface $\mathcal{S}$, DLP is equal to zero on this boundary. If we can find a Green's function of first kind on $\mathcal{S}$ then the contribution given by this boundary in the computation of $\bar{P}$ will be null. Similarly we know the normal derivative of $\bar{P}$ on the wetted region $\mathcal{P}$ then if we choose a Green's function of the second kind, DLP will be zero on this boundary. Therefore we need in this case a Green's function which verifies these boundary conditions on the boundary and which is symmetric on $(x,y,z)$ and $(\xi,\zeta,\tau)$. An appropriate Green's function, defined for $\boldsymbol{x} \in \bar{\Omega}$, is given by:
\begin{gather}
\label{eqint4}
\forall \boldsymbol{\xi} \in \bar{\Omega}, \quad G(\boldsymbol{x},\boldsymbol{\xi}) = - \frac{1}{2 \pi} \log r + \frac{1}{2 \pi} 
\log \left(\frac{\vert\boldsymbol{\xi} \vert}{R} \hat{r}\right),
\end{gather}
with $r = \vert \boldsymbol{x}-\boldsymbol{\xi} \vert = \sqrt{(x - \xi)^2 + \left(y - \zeta \right)^2}$ and $\hat{r} = \vert\boldsymbol{x}-\boldsymbol{\xi}^{*} \vert$ the distance between $\boldsymbol{x}$ and $\boldsymbol{\xi}^{*}=\frac{R^2}{\vert \boldsymbol{\xi} \vert^2} \xi$. $\boldsymbol{\xi}^{*}$ is the inverse of $\boldsymbol{\xi}$ with respect of the circle of center $(0,0)$ and radius R. Therefore by using the Green's function (\ref{eqint4}) the pressure impulse along the wet surface is given by:
\begin{gather}
\label{eqint3}
P(\boldsymbol{x}) = - 2 \int_{C_1} G(\boldsymbol{x},\boldsymbol{\xi}) 
\frac{\partial P}{\partial \zeta}(\boldsymbol{\xi}) \,\text{d} \xi
\end{gather}
\begin{figure}
\begin{center}
\includegraphics[width=7cm]{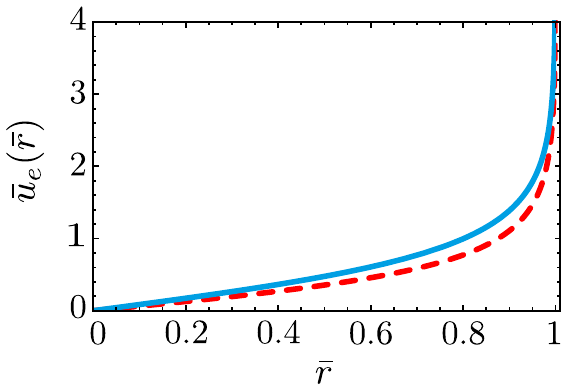}
\caption{Comparison between the slip velocity determined in the axisymmetric case (red dashed line) and in the two-dimensional case (blue continuous line).}
\label{fig:2D_axi}
\end{center}
\end{figure}
Consequently by using the substitutions $(x,y)=R (\bar{x},\bar{y})$, $(\xi,\zeta)=R (\bar{\xi},\bar{\zeta})$ and $u_{e}=U \bar{u}_{e}$, we obtain for all $\bar{\boldsymbol{x}} \in C_1$:
\begin{gather*}
\bar{u}_{e}(\bar{\boldsymbol{x}})  = - \frac{1}{\pi} \frac{\partial}{\partial \bar{x}} \int_{-1}^1 \left(-\log \left(\sqrt{(\bar{x} - \bar{\xi})^2 + \left(\bar{y} - \bar{\zeta} \right)^2}\right) \right. \\
+ \left. \log \left(\sqrt{1-2(\bar{x} \bar \xi + \bar y \bar \zeta)+(\bar{x}^2+\bar{y}^2)(\bar{\xi} ^2 +\bar{\zeta}^2)} \right)\right) \,\text{d} \bar \xi.
\end{gather*}
We finally obtain the two-dimensional slip velocity: 
\begin{gather}
\bar{u}_{e}(\bar{x})  = - \frac{1}{\pi} \left(\frac{2}{\bar x}-\left(1+ \frac{1}{\bar{x}^2}\right) \log \left(\frac{1+ \bar x}{1-\bar x}\right) \right).
\end{gather}

This last solution, contrary to the one determined in axisymmetric geometry, could be expressed explicitly. Although the formulation of these two solutions are different, their structure are very similar (see Fig.~\ref{fig:2D_axi}). As for all the previous problems studied in this paper the slip/impact velocity has a logarithmic singularity close to the contact line. In this last calculation, this singular structure directly emerge from the Green's function appropriated to this problem.



\bibliographystyle{elsarticle-num}

\bibliography{biblio}

\end{document}